\documentclass[fleqn,10pt,amsmath,amssymb]{wlscirep}
\usepackage[utf8]{inputenc}
\usepackage[T1]{fontenc}

\usepackage{epsfig,graphicx,color}
\usepackage{booktabs,siunitx}
\usepackage{float}   
\usepackage{subcaption}
\usepackage{float}
\newcommand{\nc}{\newcommand}
\usepackage{tabularx}
\newcolumntype{Y}{>{\raggedleft\arraybackslash}X}

\nc{\half}{\frac{1}{2}}
\nc{\bp}{{\bf p}}
\nc{\bpp}{{\bf p'}}
\nc{\bpz}{{\bf p''}}
\nc{\bk}{{\bf k}}
\nc{\bkp}{{\bf k'}}
\nc{\bkz}{{\bf k''}}
\nc{\bera}{\langle}
\nc{\ket}{\rangle}
\nc{\bq}{{\bf q}}
\nc{\bqp}{{\bf q'}}
\nc{\tpi}{\tilde{\pi}}
\nc{\bpi}{\boldsymbol \pi}
\nc{\btpi}{\tilde{\boldsymbol \pi}}

\title{A three-dimensional momentum-space calculation of three-body bound state in a relativistic Faddeev scheme}

\author[1,2]{M. R. Hadizadeh}
\author[3]{M. Radin}
\author[3]{K. Mohseni}
\affil[1]{College of Engineering, Science, Technology and Agriculture, Central State University, Wilberforce, OH
45384, USA,}
\affil[2]{Department of Physics and Astronomy, Ohio University, Athens, OH 45701, USA,}
\affil[3]{Department of Physics, K. N. Toosi University of Technology, Tehran, Iran.}


\begin{abstract}
In this paper, we study the relativistic effects in a three-body bound state. For this purpose, the relativistic form of the Faddeev equations is solved in momentum space as a function of the Jacobi momentum vectors without using a partial wave decomposition. The inputs for the three-dimensional Faddeev integral equation are the off-shell boost two-body $t-$matrices, which are calculated directly from the boost two-body interactions by solving the Lippmann-Schwinger equation. The matrix elements of the boost interactions are obtained from the nonrelativistic interactions by solving a nonlinear integral equation using an iterative scheme. The relativistic effects on three-body binding energy are calculated for the Malfliet-Tjon potential. Our calculations show that the relativistic effects lead to a roughly 2\% reduction in the three-body binding energy. 
The contribution of different Faddeev components in the normalization of the relativistic three-body wave function is studied in detail. 
The accuracy of our numerical solutions is tested by calculation of the expectation value of the three-body mass operator, which shows an excellent agreement with the relativistic energy eigenvalue.
\end{abstract}

\begin{document}

\flushbottom
\maketitle
\thispagestyle{empty}

\section{Introduction}
The study of relativistic effects in quantum mechanical few-body systems has been the subject of many studies of the bound 
\cite{
kondratyuk1981relativistic,
glockle1986relativistic,
kondratyuk1989relativistic,
sammarruca1992relativistic,
stadler1997relativistic,
sammarruca1998comment,
kamada2002lorentz,
carbonell2003three,
kamada2010calculations,
kamada2019derivation} 
and scattering 
\cite{kamada2000practical,
sekiguchi2005resolving,
keister2006quantitative,
witala2006elastic,
skibinski2006relativistic,
witala2011relativistic} states. 
All these studies are mainly performed using a partial wave decomposition, that leads to coupled equations for the bound or scattering amplitudes which are expanded in the angular momentum basis. 
The discrete partial wave summation of the resulting amplitudes is dependent on the energy scale of the problem, and at intermediate and high energy scales of few GeV, the convergence can be reached very slowly. 
To avoid this problem, several studies have been performed in a three-dimensional (3D) approach to formulate the bound and scattering amplitudes as a function of the Jacobi momentum vectors by replacing the discrete angular momentum quantum numbers with continuous angle variables.
While the number of mesh points for angular integrations in a 3D formalism remains constant for all energy scales, it simultaneously considers the contribution of all partial wave components. Consequently, the number of equations in a 3D scheme is independent of the energy scale of the problem \cite{elster1998two,elster1999three,schadow2000three,fachruddin2000nucleon,fachruddin2001new,fachruddin2003n,liu2003model,fachruddin2004operator,liu2005three,lin2007first,hadizadeh2007four,lin2008relativistic,lin2008poincare,bayegan2008three,bayegan2008realistic,bayegan2008low,bayegan2009three,golak2010two,glockle20103n,skibinski2010numerical,glockle2010new,glockle2010exact,hadizadeh2011solutions,skibinski2011recent,golak2013three,veerasamy2013two,hadizadeh2014three,hadizadeh2014relativistic,topolnicki2015first,radin2017four,topolnicki2017operator,topolnicki2017three}.
Thus, the 3D approach would be a powerful tool to study the scattering problems in few GeV energy scales when the partial wave formulation is much more complicated for the evaluation of permutation operators, and the numerical convergence is difficult to reach. On this basis, the 3D approach has been implemented in relativistic calculations of three-body (3B) scattering problems \cite{lin2007first,lin2008relativistic,polyzou2014relativistic,elster2007relativistic} and recently applied to the 3B bound states \cite{hadizadeh2014relativistic,hadizadeh2016relativistic}.
It is shown that the difference between the relativistic and the nonrelativistic dynamics is in i.) the momentum dependence of the kinetic energy or free propagator, ii.) in the relation of potential operators to the two-body (2B) $t-$matrices, iii.) and in the form of the momentum basis.
The last one leads to a non-identical Jacobian of the momentum variable change and also a permutation coefficient in the relativistic case. 
At the two-body level, the relativistic 2B interactions designed in such a way that the deuteron properties and 2B observables are preserved. 
The inputs for relativistic three- and four-body bound and scattering calculations are relativistic 2B $t-$matrices.
There are few options to calculate the relativistic 2B $t-$matrices including the following ones:
\begin{itemize}
\item One option is to solve the relativistic Lippmann-Schwinger equation with relativistic 2B interactions. The input relativistic interactions can be obtained directly from the nonrelativistic one by solving a quadratic equation. 
A number of computational methods have been developed to solve the nonlinear equation, including the spectral expansion method \cite{kamada2002lorentz,glockle1986relativistic} and an iterative approach \cite{kamada2007realistic}. 
In the spectral expansion method, a completeness relation of the 2B bound and scattering states is inserted into the right side of the quadratic equation and then the result is projected into the momentum space. 
The powerful iterative scheme proposed by Kamada and Gl\"ockle has been successfully implemented in two- and three-body bound and scattering state calculations using a partial wave decomposition.
In this paper, we extend the implementation of this iterative scheme to the 3B bound state calculations in a 3D scheme by calculating the relativistic 2B $t-$matrices directly from the relativistic 2B interactions.

\item The alternative option is the calculation of relativistic 2B $t-$matrices directly from the nonrelativistic one. A two-step process is used to achieve this. In the first step, the relativistic right-half-shell 2B $t-$matrices are calculated analytically from the nonrelativistic right-half-shell 2B $t-$matrices. Then the fully off-shell 2B $t-$matrices are calculated by solving the first resolvent equation with the transformation of Coester-Piper-Serduke \cite{coester1975relativistic,keister2006quantitative}. 
In this method, there is no need for the matrix elements of the relativistic 2B interactions and it has been successfully implemented in the 3B bound and scattering calculations in a 3D approach \cite{lin2007first,lin2008relativistic,polyzou2014relativistic,elster2007relativistic,hadizadeh2014relativistic,hadizadeh2016relativistic}.

\end{itemize}

The paper is organized as follows. In Sec. \ref{formalism}, a brief review to the formalism of the relativistic Faddeev equations in a three-dimensional approach is provided. 
Additionally, the formulation of the boost 2B interactions and $t-$matrices, which are essential ingredients for our calculations, are presented. 
In Sec. \ref{results} our numerical results for boost 2B interactions and $t-$matrices are given, and the contribution of different relativistic corrections to the 3B binding energy is studied. As a test for our numerical calculations, the expectation value of the 3B mass operator is calculated and compared with the relativistic binding energy.
In Sec. \ref{summary} we present our conclusions and perspectives. 

\section{Relativistic Faddeev equation for three-body bound state} \label{formalism}

The 3B mass operator of three identical particles with mass $m$ and momentum $\bk_i $, interacting with pairwise interactions, in the relativistic quantum mechanics is defined as
\begin{eqnarray} \label{eq.3N-Hamiltonian}
M=M_0+\sum_{i<j} V_{k}^{ij}.
\end{eqnarray}
The free-body mass operator $M_0$ in the 3B Hilbert space is given by
\begin{eqnarray} \label{eq.M0}
M_0=\sum_{i=1}^3 (m^2+k_i^{2})^{\half}.
\end{eqnarray}
The 2B boost interactions $V_{k}^{ij} \equiv V_k$ embedded in the 3B Hilbert space can be obtained in terms of the relativistic 2B interaction $V_r$ by following nonlinear relation
\begin{eqnarray} \label{eq.v-rel-3N}
V_{k}=\biggl ( \bigl (\omega(\hat p)+V_r \bigr)^2+ \hat k^{2} \biggr )^{\frac{1}{2} } - \biggl ( \omega^2( \hat p)+ \hat k^{2} \biggr ) ^{\half},
\end{eqnarray}
where $\bk=\bk_i+\bk_j$ is total momentum of the subsystem $(ij)$, $\omega(\hat p)=2\sqrt{m^2+\hat p^2}$, and $\bp$ is the relative momentum in the 2B subsystem $(ij)$. The $k$-dependence arises from this fact that in a 3B system, the 2B subsystems are not at rest. Obviously, for $\bk={\bf 0}$ the boost interaction reduces to the relativistic interaction, {\it i.e.} $V_{k}=V_r$. Since we work in the center of mass frame of the 3B system, $k$ can be considered as the momentum of the spectator particle. 
The Schr\"{o}dinger equation for the bound state of three identical particles interacting with pairwise forces is described by Faddeev equation
\begin{eqnarray} \label{eq.Faddev}
\psi=G_0 \, t\, P\, \psi,
\end{eqnarray}
where $G_0=(M_t-M_0)^{-1}$ is free propagator, $M_t=E_t + 3m$ is the 3B mass eigenvalue. The permutation operator is defined as $P=P_{ij}P_{jk}+P_{ik}P_{jk}$, and $t$ is 2B transition operator.
In order to solve the Faddeev equation (\ref{eq.Faddev}) in momentum space, one needs the 3B basis states, which are composed of two relativistic Jacobi momentum vectors $| \bp \, \bk \ket $, defined in details in Ref. \cite{hadizadeh2014relativistic}. The 3B Jacobi momentum variables are defined by boosting the single-particle momenta to the three-body rest frame and then boosting two-body subsystem momenta to the subsystem rest frame. 
The free-body mass operator $M_0$, given in Eq. (\ref{eq.M0}), is diagonal in these basis states
\begin{equation} \label{eq.H0-rel-diagonal}
M_0  | \bp \, \bk \ket = \biggl  ( \bigl ( \omega^2(p) + k^2 \bigr )^{\half} + \Omega(k)  \biggr  )| \bp \, \bk \ket,
\end{equation}
where $\Omega(k) = \left (m^2+k^2 \right )^{\half}$. 
As shown in Ref. \cite{hadizadeh2014relativistic} the representation of Faddeev equation (\ref{eq.Faddev}) in the relativistic basis states $| \bp \, \bk \ket $ leads to the following three-dimensional integral equation
\begin{equation} \label{eq.Faddeev-integral-final-rel}
\psi (\bp \,, \bk ) =
\frac{1}{M_t-M_0(p,k)}\int d^3 k' \, N(\bk,\bkp) \,T_{k}^{sym} \big(\bp,\btpi;\epsilon \big)
 \, \psi \bigl(\bpi,\bkp \bigr),
\end{equation}
where $\epsilon=M_t- \Omega(k)$ is 2B subsystem energy, $T_{k}^{sym}(\bp,\bpp;\epsilon)$ is symmetrized boost 2B $t$-matrix defined as $T_{k}^{sym}(\bp,\bpp;\epsilon) =T_{k}(\bp,\bpp;\epsilon) +T_{k}(\bp,-\bpp;\epsilon) $, and the Jacobian function $N(\bk,\bk')$ is defined as
\begin{equation}
N(\bk,\bk') = {\cal N}^{-1}(-\bk-\bk',\bk') \, {\cal N}^{-1}(-\bk-\bk',\bk),
\label{eq.N}
\end{equation}
where ${\cal N}$ is the square root of the Jacobian which arises from going from individual momenta $\bk$ and $\bkp$ of the subsystem to the relative momentum $\bp$ and the total 2B momentum $\bk+\bkp$ \cite{hadizadeh2014relativistic},
\begin{equation}
{\cal N}(\bk,\bk') = 
\left ( {\partial (\bk,\bk') \over 
\partial (\bp, \bk+\bk') } \right ) ^{\half} =
\left ( {\omega(p) +\omega(p) \over 
\Omega(k) +\Omega(k')}
{\Omega(k) \Omega(k') \over 
\omega(p) \omega(p)} \right ) ^{\half} .
\label{eq.jacobian-rel}
\end{equation}
In the nonrelativistic limit where the momenta are much smaller than the masses, the Jacobin function $N$ reduces to one, and consequently, the relativistic Jacobi momenta reduce to the nonrelativistic ones.
The relativistic shifted momentum arguments are defined as
\begin{eqnarray} 
\btpi &=& 
 \bk' + \frac{1}{2} C(\bk,\bk') \, \bk , \cr
\bpi &=&
 \bk + \frac{1}{2} C(\bk',\bk) \, \bk',
\label{eq.4.28}
\end{eqnarray}
where the permutation coefficients $C(\bk,\bk')$ are defined as \cite{lin2007first}
\begin{eqnarray} 
C(\bk,\bk') \equiv 1+ \frac{ \Omega(k')- 
\Omega(\vert\bk + \bk'\vert) }
{ \Omega(k') + \Omega(\vert \bk + \bk'\vert ) + 
\sqrt{\biggl ( \Omega(k') 
+ \Omega (\vert\bk + \bk' \vert) \biggr)^2-k^2} }.
\label{eq.C}
\end{eqnarray}
In the nonrelativistic limit, when the momenta are much smaller than the masses, the permutation coefficient $C(\bk,\bk')$ is equal to one. Consequently, the relativistic shifted momenta $\btpi$ and $\bpi$ reduce to the corresponding nonrelativistic ones.
As we mentioned earlier, in this paper we calculate the fully-off-shell boost 2B $t-$matrices $T_{k}(\bp,\bpp;\epsilon)$ by solving the following relativistic Lippmann-Schwinger equation
\begin{eqnarray} \label{eq.t-matrix-boost}
T_{k}(\bp,\bpp;\epsilon)=V_{k}(\bp,\bpp)+ \int d^3 p'' \, \frac{V_{k}(\bp,\bpz)}
{ \epsilon - \biggl (\omega^2(p'')+ k^2 \biggr )^{\half}}\,T_{k}(\bpz,\bpp;\epsilon).
\end{eqnarray}
To solve the three-dimensional integral equation (\ref{eq.Faddeev-integral-final-rel}), we choose a coordinate system where $\bk$ is parallel to $Z-$axis, $\bp$ is in the $X-Z$ plane, and the integration vector $\bk'$ is free in the space. 
In a spherical coordinate system the radial distance, the cosine of polar angle, and azimuthal angle for momentum vectors $\bk$, $\bp$, and $\bk'$ are given by $(k,x_k=0,\phi_k=0)$, $(p,x_p \equiv x,\phi_p=0)$, and $(k',x_{k'}  \equiv x',\phi_{k'} \equiv \phi')$, correspondingly.
Using this coordinate system, Eq.~(\ref{eq.Faddeev-integral-final-rel}) can be written explicitly as
\begin{eqnarray}
 \psi (p,k,x) &=& \frac{1}{M_t- \sqrt{\omega^2(p) + k^2 } - \Omega(k) } \, \int_{0}^{\infty} dk' k'^2
\int_{-1}^{+1} dx' \int_{0}^{2\pi} d\phi'  
\cr &\times&
 N(k,k',x') \, T^{sym}_{k} \biggl ( p,\tpi, x_{p \tpi} ; M_t-
\Omega (k) \biggr) \psi (\pi,k', x_{\pi k'}) .
 \label{eq.Faddeev_final}
 \end{eqnarray}
 where the momentum and angle variables are defined as
\begin{eqnarray}
 x &=& \hat{\bk} \cdot \hat{\bp}, \cr
x' &=& \hat{\bk} \cdot \hat{\bk}' , \cr
y &=& \hat{\bp} \cdot \hat{\bk}' = xx'+\sqrt{1-x^{2}}\sqrt{1-x'^{2}}\cos \phi' ,
 \cr
\tpi &=&\sqrt{\frac{1}{4}C^2(k,k',x') k^{2}+k'^{2}+ C(k,k',x') k k' x'} ,
 \cr 
 \pi &=&\sqrt{k^{2}+\frac{1}{4} C^2(k',k,x') k'^{2}+C(k',k,x') k k' x'},
 \cr
 x_{p \tpi}&=&\frac{\frac{1}{2}C(k,k',x')k x+k' y}{\tpi},
 \cr
x_{\pi k'} &=& \frac{k x' +\frac{1}{2} C(k',k,x') k'}{\pi} .
\label{eq.angles_momenta}
 \end{eqnarray}
The integral equation (\ref{eq.Faddeev_final}) is an eigenvalue equation in the form $\lambda \psi = K(M_t) \cdot \psi$ with the eigenvalue $\lambda=1$.
Since the kernel of the integral equation $K(M_t)$ is energy dependent, the solution of the eigenvalue equation (\ref{eq.Faddeev_final}) can be started by an initial guess for the energy $E_t$ and the search in the binding energy is stopped when $|1 - \lambda| < 10^{-6}$.

The input for the Lippmann-Schwinger integral equation (\ref{eq.t-matrix-boost}) is the 2B boost potential $V_{k}(\bp,\bpp)$ which has been traditionally calculated by the representation of Eq. (\ref{eq.v-rel-3N}) in momentum space by inserting the completeness relation of bound and scattering eigenstates of the center of mass Hamiltonian ($\omega(\hat p)+V_r$) \cite{kamada2002lorentz}. 
However, with the recent successful application of the iterative scheme proposed by Kamada and Gl\"ockle \cite{kamada2007realistic} in the calculation of the relativistic interaction from the nonrelativistic one in a 3D scheme \cite {hadizadeh2017calculation}, in this paper we utilize the same method for the calculation of the boost potential.
By following \cite{kamada2007realistic}, one can obtain the boost interaction from the nonrelativistic interaction by the following quadratic equation
\begin{eqnarray} \label{eq.v_k23-in-V}
V_{nr}=\frac{1}{4m} \biggl ( \omega_k(\hat p) \,V_{k} + V_{k}\, \omega_k(\hat{p}) + V_{k}^2 \biggr),
\end{eqnarray}
where $\omega_k(\hat p)=\biggl (\omega^2(\hat p)+ \hat k^2 \biggr)^{\half}$. 
Obviously, for $\bk={\bf 0}$ the Eq. (\ref{eq.v_k23-in-V}) can be reduced to the connection between the relativistic and nonrelativistic interactions as
\begin{eqnarray} \label{eq.V-v}
V_{nr}=\frac{1}{4m} \biggl(\omega(\hat p)V_r+ V_r \omega(\hat p)+V_r^2 \biggr).
\end{eqnarray}
Equations (\ref{eq.V-v}) and (\ref{eq.v_k23-in-V}) have same operator forms, where $V_r$ and $\omega(\hat p)$ are replaced by $V_{k}$ and $\omega_k(\hat p)$.
So the procedure of calculation of boost potential $V_k$ from the nonrelativistic potential $V_{nr}$ in Eq. (\ref{eq.v_k23-in-V}) is similar to the calculation of the relativistic potential $V_r$ from the nonrelativistic potential $V_{nr}$ in Eq. (\ref{eq.V-v}).
So we can obtain the boost potential matrix elements by taking the nonrelativistic potential and by solving the following three-dimensional integral equation 
\begin{eqnarray} \label{eq.v_k23-matrix}
V_{k} (\bp,\bpp) + \frac{1}{\omega_k(p)+\omega_k(p')} \int d^3p'' V_{k} (\bp,\bpz) \, V_{k} (\bpz,\bpp) =\frac{4m \, V_{nr}(\bp,\bpp) }{\omega_k(p)+\omega_k(p')}.
\end{eqnarray}
The integral equation (\ref{eq.v_k23-matrix}) is solved using an iterative scheme and the details of the calculations are similar to the ones discussed in Ref. \cite{hadizadeh2017calculation}. We would like to mention that using an averaging scheme on two successive iterations leads to a faster convergence in the calculation of the boost interaction.

\section{Numerical Results} \label{results}

\begin{figure}[H]
\centering
  \includegraphics[width=0.6\columnwidth]{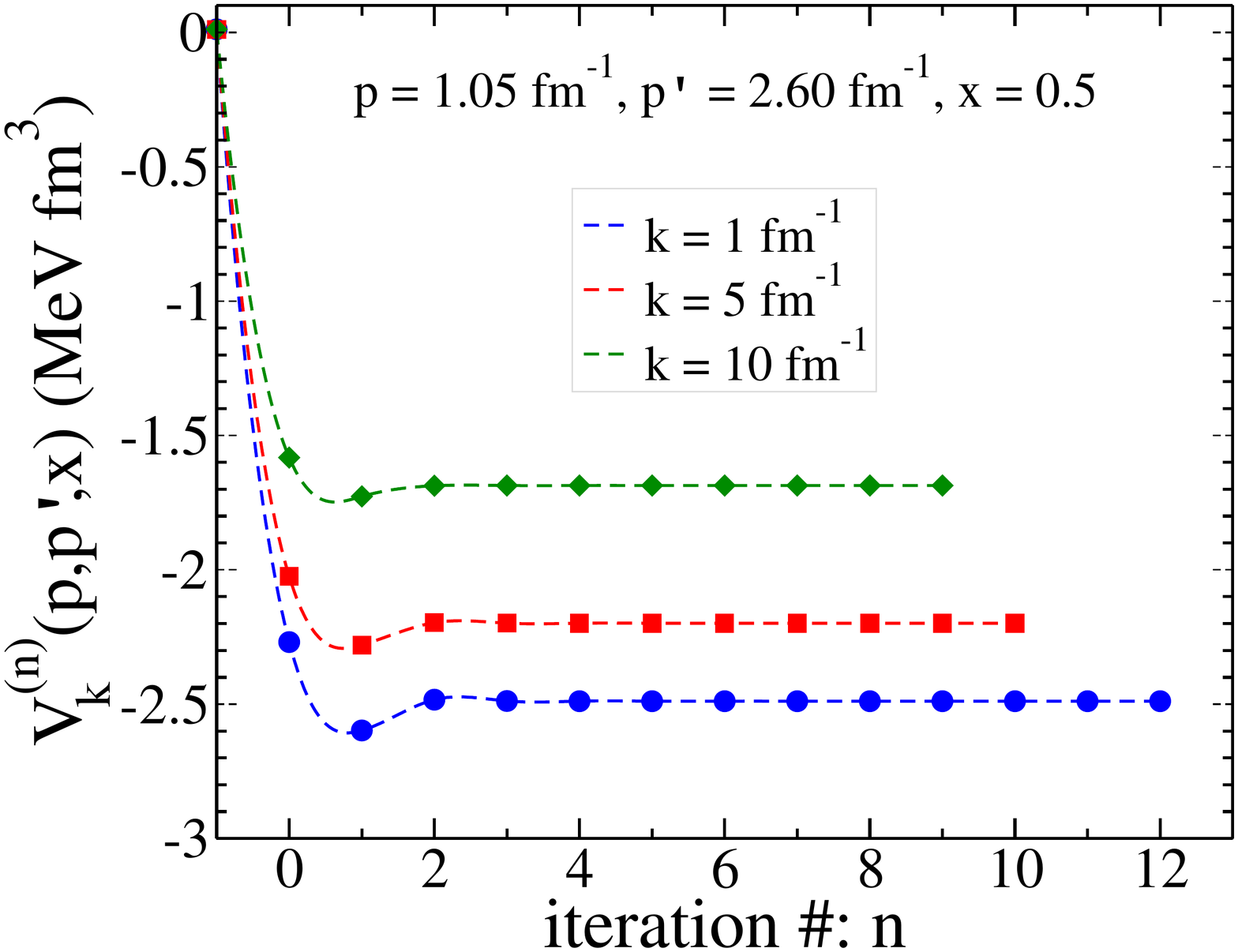}
  \caption{The convergence of the matrix elements of the boost potential $V_k(p,p',x')$ as a function of the iteration number calculated for the MT-V bare potential in the fixed points ($p=1.05$ fm$^{-1}$, $p'=2.60$ fm$^{-1}$, $x'=0.50$) with $k=1, \, 5, 10\,\, \text{fm}^{-1}$. The left endpoint of each plot is the value of the nonrelativistic MT-V potential $V_{nr}(p,p',x') = 1.1579249 \cdot 10^{-2} \, \text{MeV\,fm}^{3}$.}
  \label{fig_convergence}
\end{figure}

\begin{figure}[H]
\centering
\begin{subfigure}[t]{.3\textwidth}
\centering
\includegraphics[width=\linewidth]{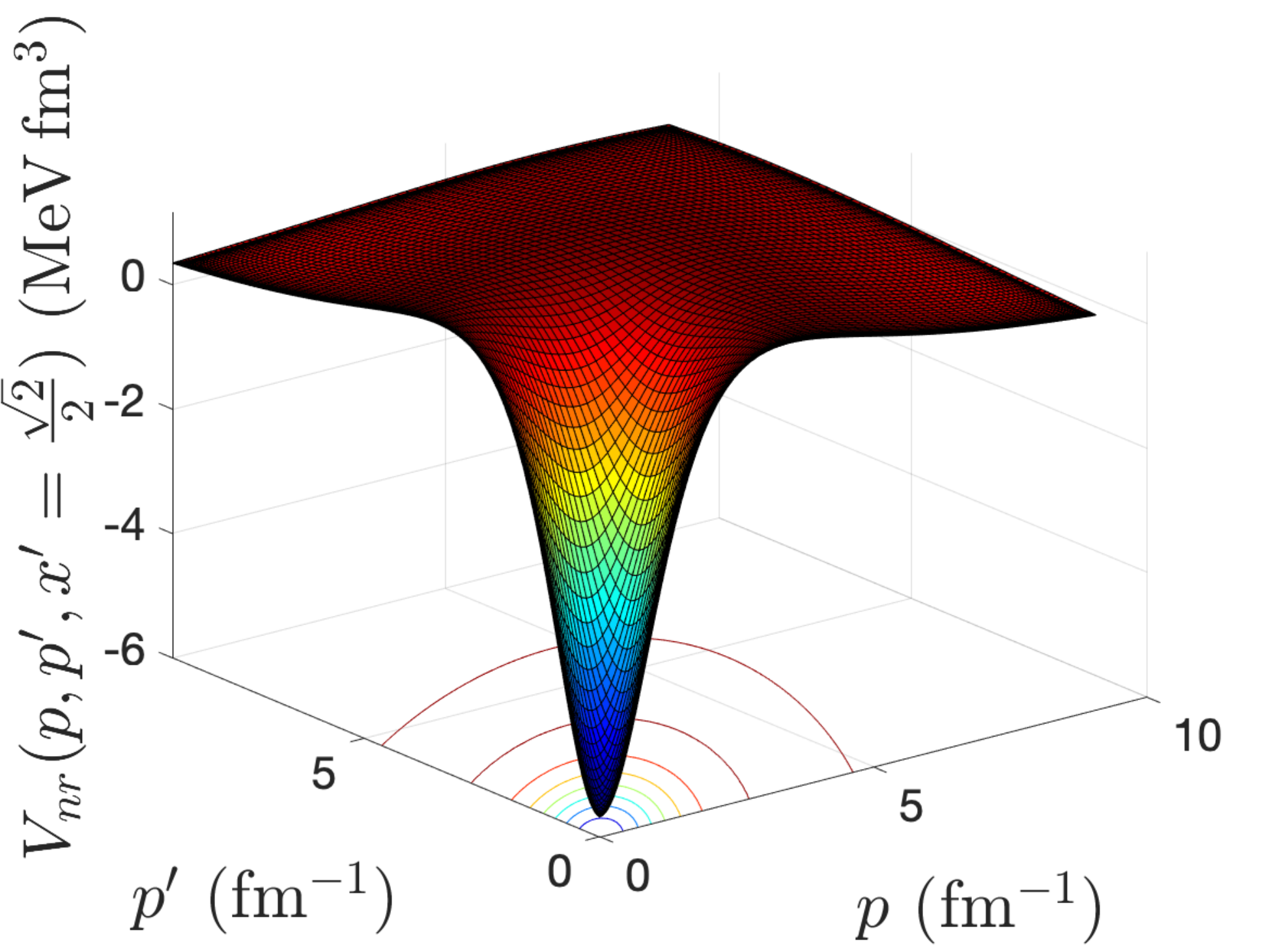}
    \caption{}\label{fig:fig_a}
\end{subfigure}
\begin{subfigure}[t]{.3\textwidth}
\includegraphics[width=\linewidth]{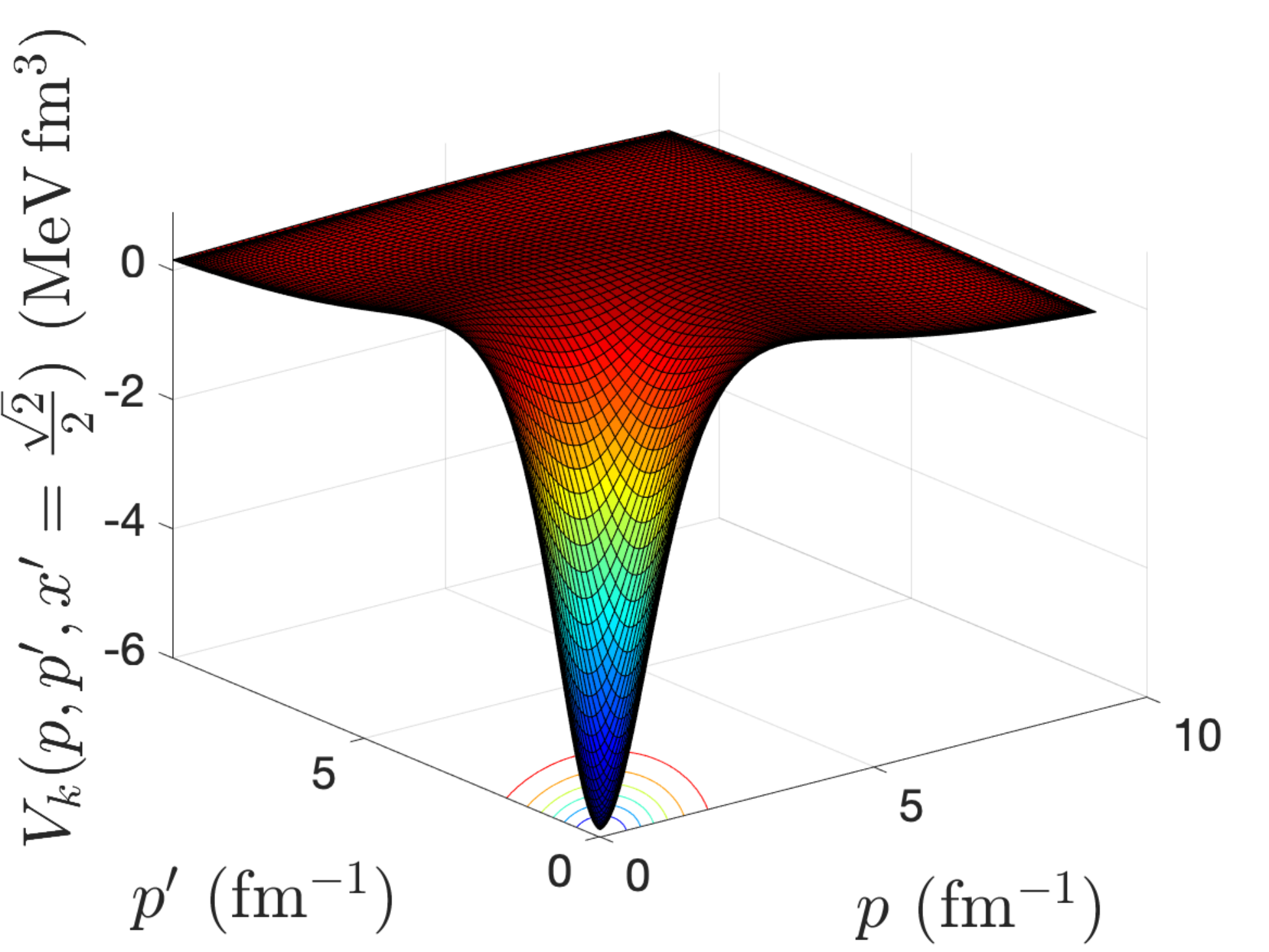}
\caption{}\label{fig:fig_b}
\end{subfigure}
\begin{subfigure}[t]{.3\textwidth}
\centering
\includegraphics[width=\linewidth]{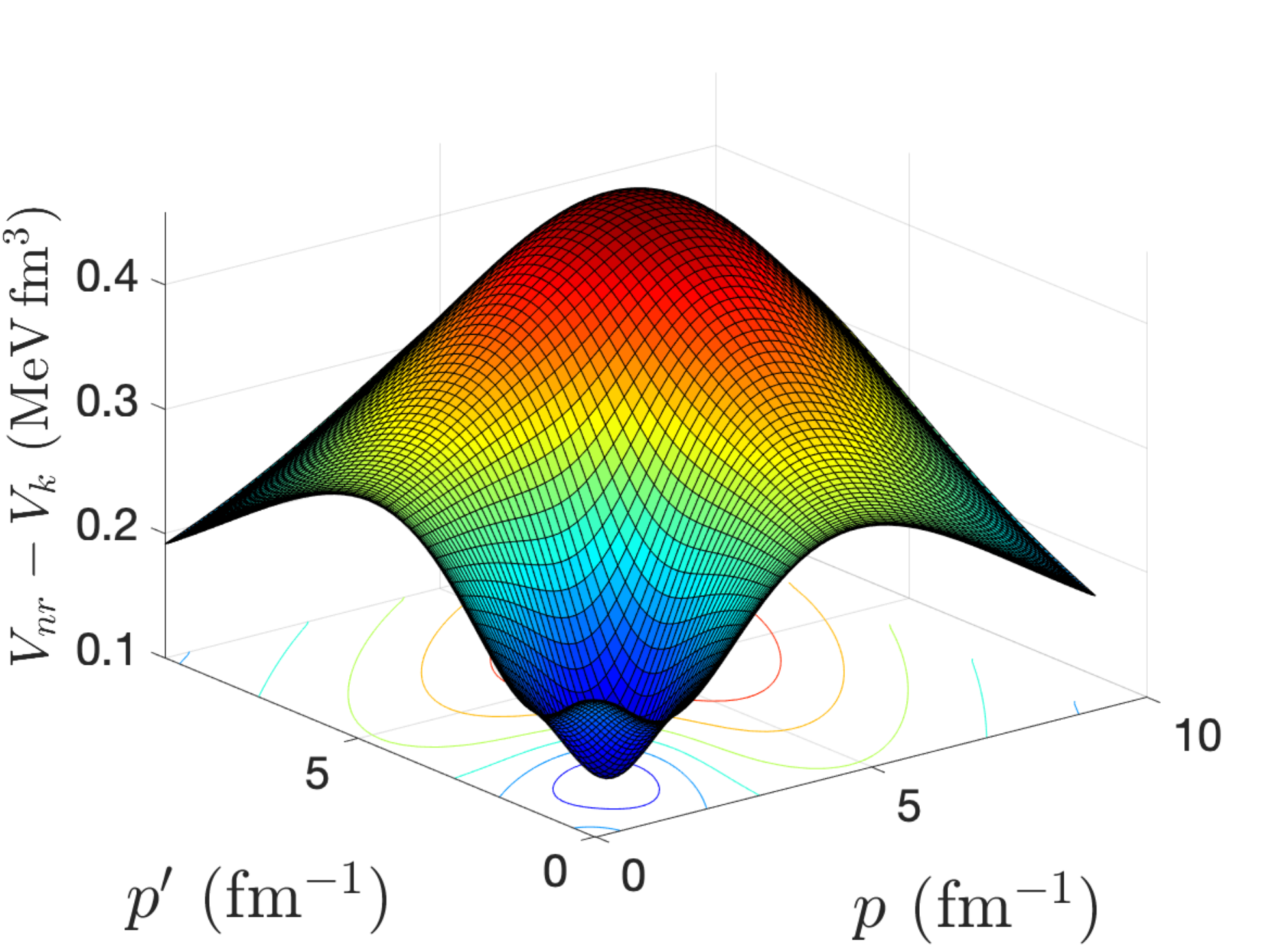}
    \caption{}\label{fig:fig_c}
\end{subfigure}
\begin{subfigure}[t]{.3\textwidth}
\centering
\includegraphics[width=\linewidth]{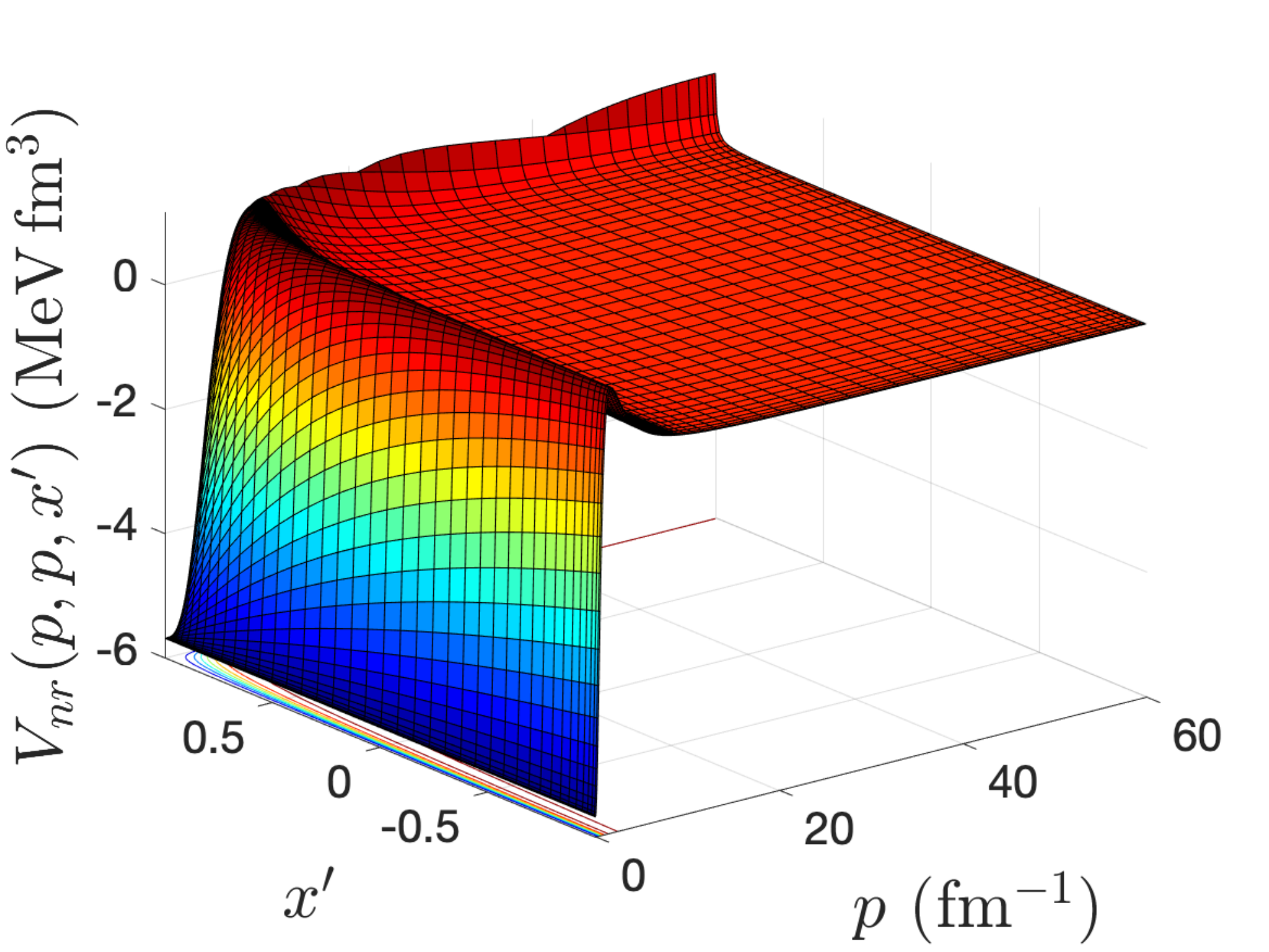}
    \caption{}\label{fig:fig_d}
\end{subfigure}
\begin{subfigure}[t]{.3\textwidth}
\includegraphics[width=\linewidth]{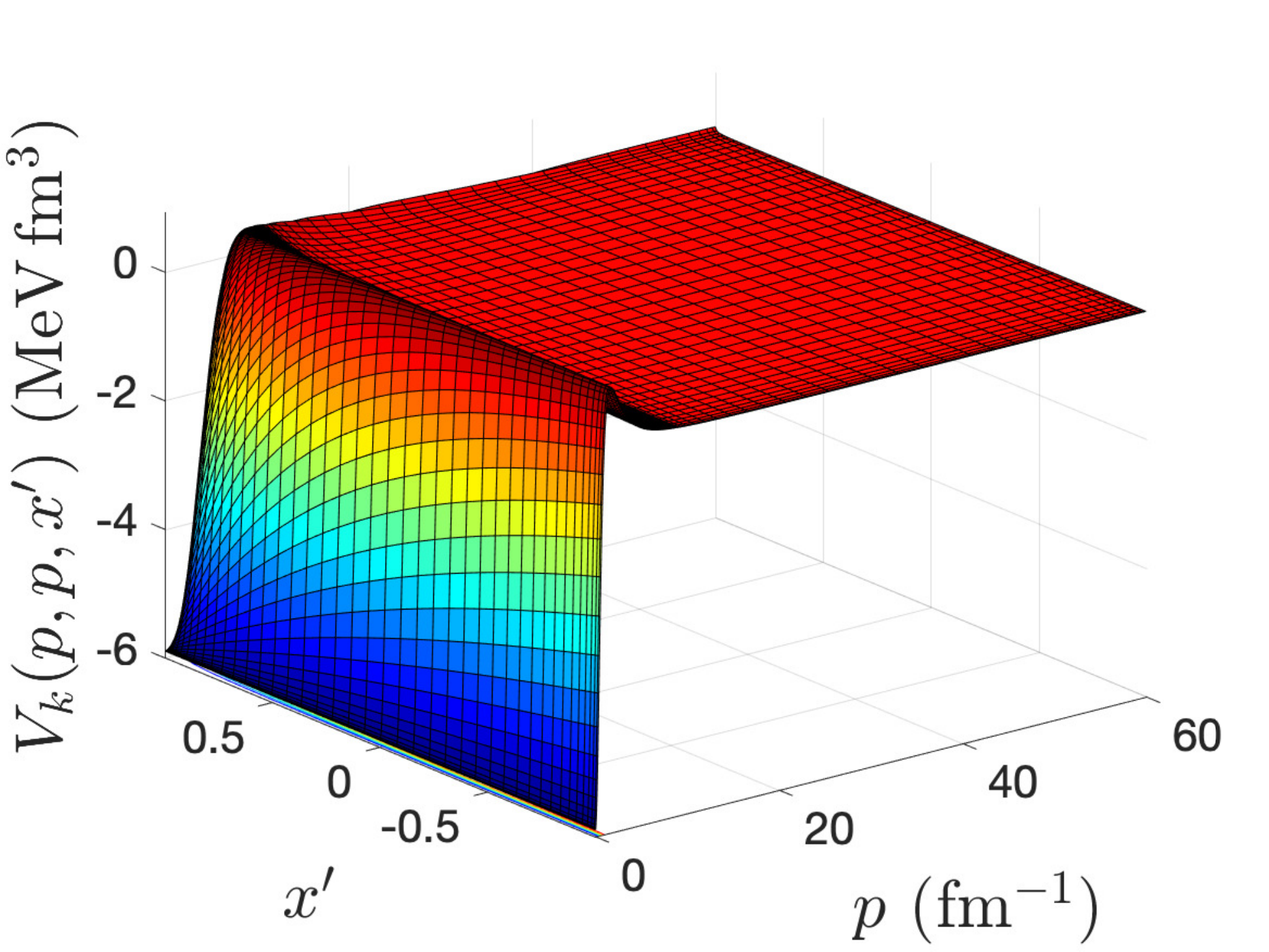}
\caption{}\label{fig:fig_e}
\end{subfigure}
\begin{subfigure}[t]{.3\textwidth}
\centering
\includegraphics[width=\linewidth]{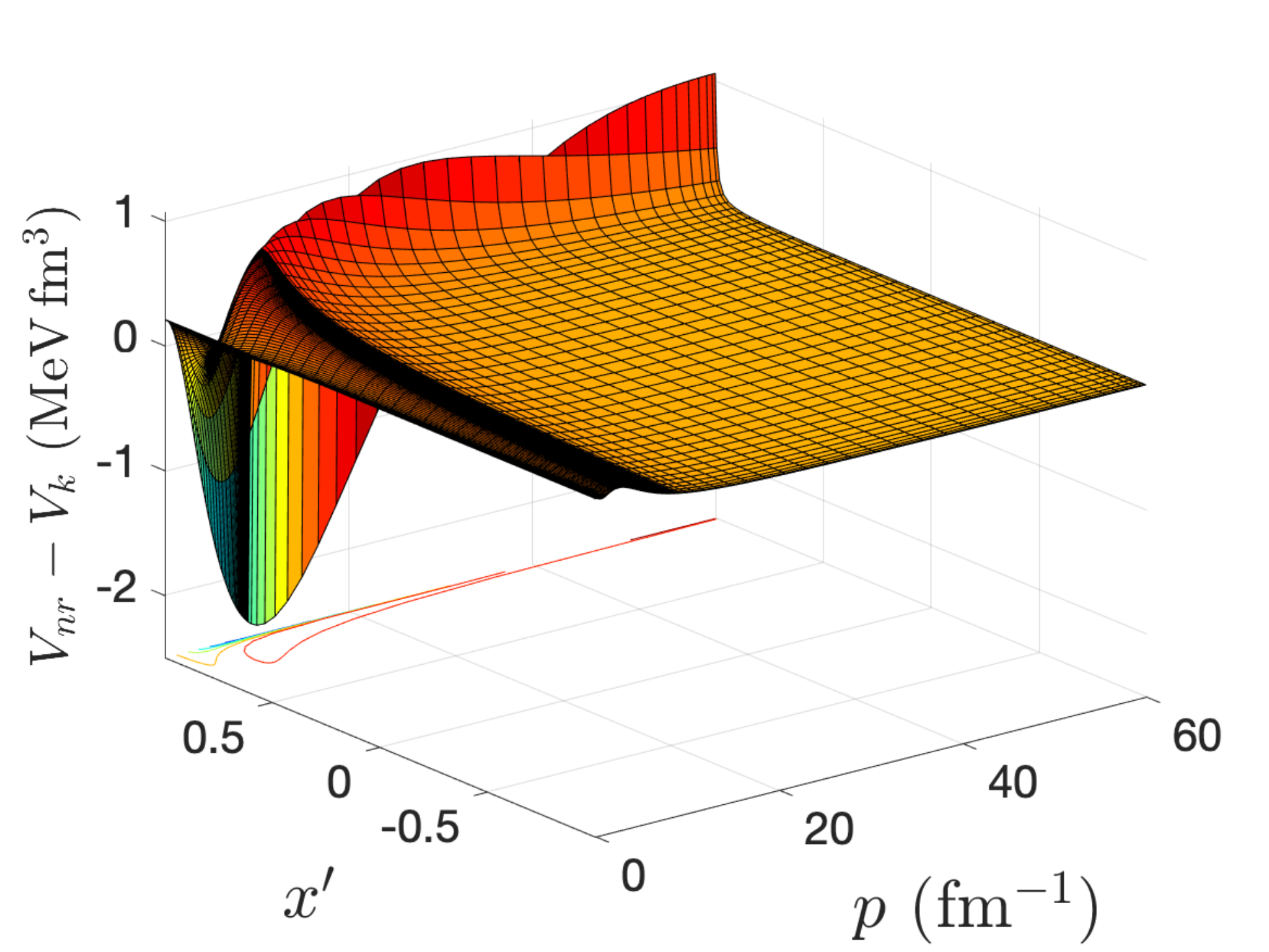}
    \caption{}\label{fig:fig_f}
\end{subfigure}
\caption{
The matrix elements of the nonrelativistic (a \& d), the boost (b \& e) 2B potentials and their differences (c \& f) calculated with the MT-V potential. They are shown as a function of the 2B relative momenta $p$ and $p'$ for the angle $x'=\frac{\sqrt{2}}{2}$ (upper panel) and as a function of the 2B relative momenta $p=p'$ and the angle between them $x'$ (lower panel). The boost potential is obtained with $k=1\, \text{fm}^{-1}$.}
  \label{fig+V_pp_px}
\end{figure}

In the first step we solve the three-dimensional integral equation (\ref{eq.v_k23-matrix}) to calculate the matrix elements of the boost interaction $V_{k}(p,p',x')$ from the nonrelativistic interaction $V_{nr}(p,p',x')$ by an iterative scheme. The iteration starts with an initial guess $V_{k}^{(0)}(p,p',x')=\frac{4m \, V_{nr}(p,p',x') }{\omega_k(p)+\omega_k(p')}$ and then continued to reach a convergence in the matrix elements of the boost potential with a relative error of $10^{-6}$ at each set of points $(p,p',x')$. In Table \ref{table:MT-V-convergence1} we show an example of the convergence of the matrix elements of the boost potential $V_k(p,p',x')$ as a function of the iteration number calculated for three different values of the Jacobi momentum $k=1, \, 5, \, 10\, \text{fm}^{-1}$ for the Malfliet-Tjon-V (MT-V) bare potential in the fixed points ($p=1.05$ fm$^{-1}$, $p'=2.60$ fm$^{-1}$, $x'=0.50$). As one can see for larger values of the Jacobi momentum $k$ the convergence is reached faster. The data of Table \ref{table:MT-V-convergence1} are also illustrated in Fig. \ref{fig_convergence}. In Fig. \ref{fig+V_pp_px} the matrix elements of the nonrelativistic and the boost interactions and also the difference between them are shown.

\begin{figure}[H]
\centering
  \includegraphics[width=.7\columnwidth]{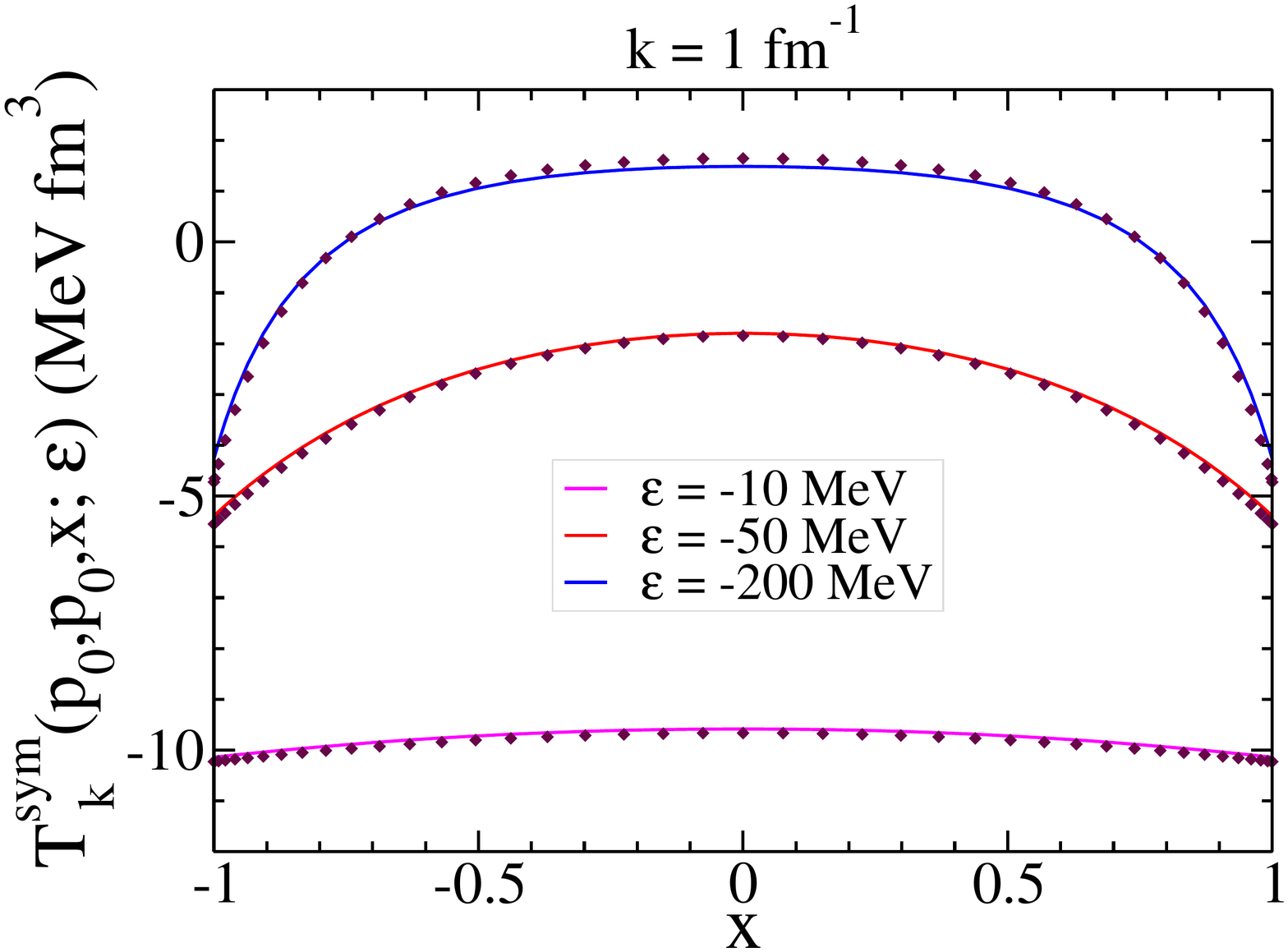}
  \caption{The angular dependency of the symmetrized boost 2B $t-$matrix $T_k^{sym}(p_0,p_0,x,\epsilon)$ calculated for $k=1 \, \text{fm}^{-1}$ and energies $\epsilon=-10, -50, -200$ MeV, with $p_0=\sqrt{m\cdot |\epsilon|}$. The diamond symbols indicate the corresponding nonrelativistic $t-$matrices. The input boost potential is obtained from the MT-V potential.}
  \label{fig_T_onshell}
\end{figure}

By having the matrix elements of the boost interactions $V_{k}(p,p',x')$ we solve the Lippmann-Schwinger integral equation (\ref{eq.t-matrix-boost}) to calculate the off-shell boost $t-$matrices $T_{k}(p,p',x';\epsilon)$ and then symmetrize it on the angle variable to get $T_{k}^{sym}(p,p',x';\epsilon)$. In Fig. \ref{fig_T_onshell} we show the angular dependencies of the symmetrized 2B boost $t-$matrices for energies $\epsilon=-10, -50, -200\, \text{MeV}$ obtained for $k=1\, \text{fm}^{-1}$ and compared with the corresponding nonrelativistic $t-$matrices. As we expect for higher energies the difference between the boost and the nonrelativistic $t-$matrices is more visible.

\begin{table}[H]
\centering
\begin{tabular}{cccc}
\hline
 & \multicolumn{3}{c}{$V_{nr}(p,p',x')$} \\ \cline{2-4}
 & \multicolumn{3}{c}{$1.1579249 \cdot 10^{-2}$}
  \\ 
  \hline
  & \multicolumn{3}{c}{$V_{k}(p,p',x')$} \\ \cline{2-4}
 Iteration \# & $k=1 \, \text{fm}^{-1}$ & $k=5\, \text{fm}^{-1}$ & $k=10\, \text{fm}^{-1}$  \\ 
 \hline
      0 &  -2.2688550    & -2.0244444 & -1.5822847 \\
      1 &  -2.5977481    & -2.2796733 & -1.7267466  \\
      2 &  -2.4833116    & -2.1967138 & -1.6871314  \\
      3 &  -2.4880714    & -2.1984435 & -1.6861455  \\
      4 &  -2.4886996    &  -2.1988384 & -1.6861841  \\
      5 &  -2.4887497    &  -2.1988784 & -1.6861961  \\
      6 &  -2.4887515    &  -2.1988795 & -1.6861962  \\
      7 &  -2.4887519    &  -2.1988792 & -1.6861957  \\
      8 &  -2.4887523    &  -2.1988792 & -1.6861954  \\
      9 &  -2.4887525    &  -2.1988793  & -1.6861954  \\
     10 &  -2.4887526    &  -2.1988793 & --  \\
     11 &  -2.4887527    &  -- &-- \\
     12 &  -2.4887527   & -- & -- \\

\hline
\end{tabular}
\caption{The convergence of the matrix elements of the boost potential $V_k(p,p',x')$ (in units of MeV fm$^3$) as a function of the iteration number calculated using the MT-V bare potential in the fixed points ($p=1.05$ fm$^{-1}$, $p'=2.60$ fm$^{-1}$, $x'=0.50$) and for three different values of the Jacobi momentum $k=1, \, 5, \, 10\, \text{fm}^{-1}$. The value of the MT-V bare potential $V_{nr}(p,p',x')$ is also given.} \label{table:MT-V-convergence1}
\end{table}

For the calculation of relativistic effects in the 3B binding energy, we solve the three-dimensional integral equation (\ref{eq.Faddeev_final}). To this aim, for discretization of continuous momentum and angle variables, we use Gauss-Legendre quadrature. For angle variables, a linear mapping, and for Jacobi momenta, a hyperbolic-linear mapping is used. The momentum cutoffs used for Jacobi momenta $p$ and $k$ are $60\, \text{fm}^{-1}$ and $20\, \text{fm}^{-1}$, correspondingly. In Table \ref{table:MT-V_triton}, we present our numerical results for the relativistic 3B binding energy as a function of the number of mesh points for the Jacobi momenta $p$ and $k$. For discretization of polar and azimuthal angles, we use 40 mesh points. The symmetry property of the Faddeev component on the angle $x$, {\it i.e.} $\psi(p,k,x) = \psi(p,k,-x)$, is employed to save memory and computational time in solving Eq. (\ref{eq.Faddeev_final}). As one can see, the relativistic 3B binding energy for the MT-V potential converges to $-7.5852$ MeV. 

\begin{table}[H]
\begin{center}
\begin{tabular}{ccc }
\hline
 $N_p$ &  $N_k$ & $E^{3B}_{r}$ \\ \hline
20 &  20 &  -7.8056  \\
30 &  30 &   -7.6448  \\
40 &  40 &   -7.5958  \\
50 &  50 &   -7.5956 \\
60 &  60 &   -7.5883  \\
70 &  70 &   -7.5863  \\
80 &  80 &   -7.5856  \\
90 &  90 &   -7.5853  \\
100 & 100 &  -7.5853   \\
110 & 110 &   -7.5852  \\
120 & 120 &   -7.5852  \\
130 & 130 &   -7.5852  \\
 \hline
\end{tabular}
\end{center}
\caption{The convergence of the relativistic 3B binding energy (in units of MeV) as a function of the number of mesh points for Jacobi momenta $p$ and $k$. The number of mesh points for polar and azimuthal angles is 40. The results are obtained for the MT-V potential.}
\label{table:MT-V_triton}
\end{table}

In Table \ref{table:contribution}, we present the contribution of different relativistic corrections to the 3B binding energy. While the Faddeev calculations lead to the nonrelativistic 3B binding energy of $-7.7380$ MeV, the contribution of different relativistic corrections in the 3B binding energy is as
\begin{itemize}
\item the Jacobian function $N$ decreases the 3B binding energy with about $0.06$ MeV,
\item the permutation coefficient $C$ increases the 3B binding energy with about $0.05$ MeV,
\item the relativistic 3B free propagator $G_0$ increases the 3B binding energy with about $0.18$ MeV,
\item the relativistic 2B $t-$matrices decreases the 3B binding energy with about $0.28$ MeV.
\end{itemize}
So, two of the relativistic corrections, $N$ and 2B $t-$matrices, decrease the 3B binding energies, and the other two terms, $C$ and $G_0$, increase the 3B binding energy. When we apply all the corrections together, we obtain a relativistic 3B energy of $-7.5852$ MeV, which indicates a reduction of about $-0.15$ MeV or a percentage difference of $2\%$.

As we discussed earlier, an alternative approach for the calculation of the boost 2B $t-$matrices is solving a first resolvent equation, where the boost 2B $t-$matrices can be obtained directly from the nonrelativistic 2B $t-$matrices. In this approach, there is no need for the matrix elements of relativistic or boost potentials defined in Eqs. (\ref{eq.v_k23-in-V}) and (\ref{eq.V-v}). The implementation of this method to 3B bound state calculations leads to a reduction of about 3.3\% in the 3B binding energy for the MT-V potential \cite{hadizadeh2014relativistic}. 
As one can see, the calculation of the boost 2B $t-$matrices using two different methods, {\it i.e.} solving the first resolvent equation discussed in Ref. \cite{hadizadeh2014relativistic} and the solution of the relativistic Lippmann-Schwinger equation given in Eq. (\ref{eq.t-matrix-boost}), leads to a slightly different correction in the 3B binding energy. Solving the relativistic Lippmann-Schwinger equation using the boost potential of Eq. (\ref{eq.v_k23-matrix}), as discussed in this paper, leads to a smaller reduction in the 3B binding energy, whereas the solution of the first resolvent equation leads to a more reduction, about 1.3\% larger.
A comparison with the results of Gl\"ockle et al. in Ref. \cite{glockle1986relativistic} for an $s-$wave projection of MT-V potential indicates a reduction of about 2.7\%, whereas the relativistic corrections for different models of 2B interactions by Kamada {\it et al.} in Ref. \cite{kamada2002lorentz} shows a wide range of decrease between 2 to 6\%.

\begin{table}[H]
\begin{center}
\begin{tabular}{ccccccc}
\hline
Relativistic correction  && $E^{3B}$ && value in MeV && $\left (\frac{E^{3B}_{app} - E^{3B}_{nr}}{E^{3B}_{nr}} \cdot 100 \right ) \%$ \\ \hline
-- && $E^{3B}_{nr}$ && $-7.7380$ && -- \\ \hline
$N$ (the Jacobian function defined in Eq. \ref{eq.N}) && $E^{3B}_{app}$ && $-7.6748$ && -0.8167 \\ \hline
$C$ (the permutation coefficients $C$ defined in Eq. \ref{eq.C}) && $E^{3B}_{app}$  && $-7.7900$ && +0.6720  \\ \hline
$G_0^{rel}$ (the relativistic 3B free propagator) && $E^{3B}_{app}$ && $-7.9183$ && +2.3301 \\ \hline
$T_{k}$ (the boost 2B $t-$matrix given in Eq. \ref{eq.t-matrix-boost}) && $E^{3B}_{app}$  && $-7.4540$ && -3.6702 \\ \hline
Full relativistic ($N$ + $C$ + $G_0^{rel}$ + $T_{k}$) && $E^{3B}_{r}$  && $-7.5852$  && -1.9747 \\
 \hline
\end{tabular}
\end{center}
\caption{The contributions of different relativistic corrections to the 3B binding energy calculated for the MT-V potential.}
\label{table:contribution}
\end{table}

By solving the three-dimensional integral equation (\ref{eq.Faddeev_final}) and obtaining the 3B binding energy and the Faddeev component $\psi (p,k,x)$ one can calculate the 3B wave function $\Psi (p,k,x)$ by adding up the Faddeev components in three different 3B clusters. The details of the calculation are addressed in Ref. \cite{hadizadeh2014relativistic} and here we briefly represent the explicit form of the 3B wave function as
\begin{equation}
\Psi (p,k,x) = \sum_{i=1}^3 \psi_i (p,k,x),
 \label{eq.Psi_sum}
\end{equation}
where the Faddeev components $\psi_1, \psi_2, \psi_3$, corresponding to the 3B clusters $(23,1), (31,2), (12,3)$, are given by
\begin{eqnarray}
\psi_1 (p,k,x) &=& \psi (p,k,x),  \cr  
\psi_2 (p,k,x) &=& \dfrac{{\cal N}(\bk_2,\bk_3)}{{\cal N}(\bk,\bk_3)}  \psi (p_2,k_2,x_2) , \cr 
\psi_3 (p,k,x) &=& \dfrac{{\cal N}(\bk_2,\bk_3)}{{\cal N}(\bk,\bk_2)}  \psi (p_3,k_3,x_3).
 \label{eq.psi_components}
\end{eqnarray}
The Jacobian function ${\cal N}$ is defined in Eq. (\ref{eq.jacobian-rel}) and the shifted momentum arguments $p_2, k_2, p_3, k_3$ and the angle variables $x_2, x_3$ are defined as
\begin{eqnarray}
 p_2 &=&  \left | \xi_p \bp +  \xi_k \bk \right | = \sqrt { \xi_p^2 \, p^2 + \xi_k^2 \, k^2 + 2\, \xi_p \, \xi_k\, p \,k \, x }, 
 \cr
k_2 &=&   \left |  \bp + \alpha \bk  \right |  =  \sqrt{  p^2 +\alpha^2 k^2 +2 \, \alpha \, p \, k \, x  }, \cr
x_2 &\equiv& \hat\bp_2\cdot \hat\bk_2 =   \dfrac{ \xi_p\, p^2 + \alpha \, \xi_k \, k^2  + (\alpha \, \xi_p + \xi_k) p \, k \, x }{ p_2 \, k_2}, 
\cr
p_3 &=& \left | \gamma_p \bp +  \gamma_k \bk \right | = \sqrt { \gamma_p^2 \, p^2 + \gamma_k^2 \, k^2 + 2\, \gamma_p \, \gamma_k\, p \,k \, x }, 
 \cr
k_3 &=&  \left | - \bp - \beta \bk  \right |  =  \sqrt{  p^2 +\beta^2 k^2 +2 \, \beta \, p \, k \, x  }, \cr
x_3 &\equiv& \hat\bp_3  \cdot \hat\bk_3 = -  \dfrac{ \gamma_p\, p^2 + \beta \, \gamma_k \, k^2  + (\beta \, \gamma_p + \gamma_k) p \, k \, x }{ p_3 \, k_3},
 \label{eq.4.47}
\end{eqnarray}
where
\begin{eqnarray}
  \alpha &=& \dfrac{1}{\omega(p)} \left (  \dfrac{p k x  }{ \omega(p) + \sqrt{ \omega(p)^2 + k^2} } - \dfrac{1}{2} \omega(p)  \right), \cr
  \beta &=& 1+ \alpha , \cr
  \gamma_p &=& \dfrac{ \left (  \dfrac{ p k x + \beta k^2  }{ \sqrt{\biggl(\Omega(k_1)+\Omega(k_2) \biggr)^2-k_3^2} + \Omega(k_1)+\Omega(k_2) } - \Omega(k_1) \right)}{\sqrt{\biggl(\Omega(k_1)+\Omega(k_2) \biggr)^2-k_3^2}} 
 , \cr 
  \gamma_k &=&    1+ \gamma_p  \beta , \cr
  \xi_p &=&  -1 +  
  \dfrac{\left (  \dfrac{ -p^2 - \alpha \beta k^2 - ( \alpha + \beta ) p k x }{ \sqrt{\biggl(\Omega(k_1)+\Omega(k_3) \biggr)^2-k_2^2} + \Omega(k_1)+\Omega(k_3) } + \Omega(k_3) \right)}{\sqrt{\biggl(\Omega(k_1)+\Omega(k_3) \biggr)^2-k_2^2}} 
  , \cr 
  \xi_k &=&   \alpha ( \xi_p+1 )  -\beta .
  \label{eq.4.45}
\end{eqnarray}
The 3B wave function is normalized as
\begin{eqnarray}
\bera \Psi  | \Psi\ket =  \sum_{i,j=1}^3 \bera \psi_i  | \psi_j \ket = 
8 \pi^2 \int_0^{\infty} dp \, p^2 
\int_0^{\infty} dk \, k^2 \int_{-1}^{+1} dx \,
\Psi^2(p,k,x)  = 1,
\label{eq.normalization_3B}
\end{eqnarray}
and the normalization of the Faddeev components is chosen as
\begin{eqnarray}
| \psi_i \ket &=&  \dfrac{| \psi_i \ket}{ \sum_{j=1}^3 \bera  \psi_j | \Psi\ket } , i=1,2,3.
\label{eq.Faddeev_normalization}
\end{eqnarray}
To study the contribution of different Faddeev components in the normalization of the 3B wave function, we show the inner product of the Faddeev components, {\it i.e.} $\bera \psi_i  | \psi_j \ket $, in Table \ref{table:product}. 
The calculated Faddeev components defined in Eq. (\ref{eq.Faddeev_normalization}) satisfy the normalization of the 3B wave function given in Eq. (\ref{eq.normalization_3B}) with an error of the order of $10^{-12}$.
While three Faddeev components defined in Eq. (\ref{eq.psi_components}) have completely different momentum dependence, 
they are all normalized to almost the same value, {\it i.e.} $\bera \psi_i  | \psi_i \ket = 0.12658$. Moreover, the inner products of the nonidentical Faddeev components leads to the same value of $\bera \psi_i  | \psi_j \ket = 0.10338$. 
Our numerical analysis confirms the normalization condition of the 3B wave function $\bera \Psi | \Psi\ket = 3 \bera \psi | \psi\ket + 3 \bera \psi | P| \psi\ket = 1$ with an accuracy of at least five significant digits.
\begin{table}[H]
\begin{center}
\begin{tabular}{cccc}
\hline
  & $j=1$ & $j=2$ & $j=3$ \\ 
 \hline
 $i=1$ & 0.12657 & 0.10338 & 0.10338 \\ 
 \hline
 $i=2$ & 0.10338 & 0.12658  & 0.10338 \\
  \hline
 $i=3$ & 0.10338 & 0.10338 & 0.12658 \\   
 \hline
\end{tabular}
\end{center}
\caption{The inner product of the Faddeev components $\bera \psi_i   | \psi_j \ket $ and their contribution in the normalization of the 3B wave function $|\Psi \ket$ defined in Eq. (\ref{eq.normalization_3B}). }
\label{table:product}
\end{table}
As a test for the numerical accuracy of our calculations, we calculate the expectation value of the 3B mass operator $\langle \Psi |M| \Psi \rangle$ and compared it with the 3B binding energy obtained from the solution of the integral equation (\ref{eq.Faddeev_final}).
By considering the definition of the 3B mass operator $M$ in Eq. (\ref{eq.3N-Hamiltonian}), the expectation value of $\langle M \rangle$ can be obtained as
\begin{equation}
\langle \Psi |M| \Psi \rangle = \langle \Psi |M_0| \Psi \rangle
   +  \langle \Psi | \sum_{i<j} V_{k}^{ij} | \Psi \rangle ,
    \label{eq:expectation}
\end{equation}
where the expectation values of the free-body mass operator and the boost potential are given by
\begin{equation}
\langle \Psi |M_0| \Psi \rangle = 8 \pi^2 \int_0^{\infty}
dp \, p^2 \int_0^{\infty} dk \, k^2  
\left ( \Omega(k) + \sqrt{\omega^2(p) + k^2} \right ) 
\int_{-1}^1 dx \, \Psi^2(p,k,x)  , \label{eq:expectation_H0}
\end{equation}
\begin{eqnarray}
\langle \Psi | \sum_{i<j} V_{k}^{ij} | \Psi \rangle &\equiv& 3 \langle \Psi | V_{k} | \Psi \rangle \cr
&=&
3 \cdot 8 \pi^2 \int_0^{\infty}
 dp \, p^2 \int_0^{\infty} dk \, k^2 \int_0^{\infty}
 dp' \, p'^2 \int_{-1}^1 dx \int_{-1}^1 dx'
 \, \Psi(p,k,x) \, v_k(p,p',x,x') \,
 \Psi(p',k,x'), \label{eq:expectation_V}
\end{eqnarray}
where $v_k(p,p',x,x')$ is the result of the azimuthal angle integration on the matrix elements of the boost potential defined as
\begin{equation}
v_k(p,p',x,x') = \int_0^{2\pi} d\phi \;
V_k(p,p',y=xx'+\sqrt{1-x^2} \, \sqrt{1-x'^2}\cos \phi).
 \label{eq:phi_integration}
\end{equation}
In Table \ref{table:MT_expectation}, we present our numerical results for the expectation value of the 3B mass operator compared with the relativistic 3B binding energy. We also show our results for the expectation value of the nonrelativistic 3B Hamiltonian and the nonrelativistic 3B binding energy.
As one can see the expectation value of free-body mass operator is $\langle M_0 \rangle = 3m + 28.3876\, \text{MeV}$, while the expectation value of the boost potential is $\langle V_k \rangle = -35.9716\, \text{MeV}$ which leads to the expectation value of the 3B mass operator of $\langle M \rangle = 3m -7.5840 \, \text{MeV}$. The difference between the relativistic 3B binding energy, obtained from three-dimensional Faddeev integral equation (\ref{eq.Faddeev_final}), and the expectation value of the 3B mass operator, obtained from Eq. (\ref{eq:expectation}), is about $0.0012\, \text{MeV}$. Our numerical results for the nonrelativistic solution of the three-dimensional Faddeev integral equation leads to the binding energy of $E_{nr}^{3B} = -7.7380 \, \text{MeV}$ and the nonrelativistic expectation value of $\langle H \rangle = -7.7384 \, \text{MeV}$ with a difference of $0.0004 \, \text{MeV}$.
As one can see for both relativistic and nonrelativistic calculations, the calculated 3B binding energy is in excellent agreement with the expectation value of the 3B Hamiltonian.

\begin{table}[H]
\begin{center}
\begin{tabular}{cccccccccccccc}
\hline
 && $\langle M_0 \rangle - 3m$ &&  $\langle V_k \rangle$ && $\langle M \rangle -3m $ && $E_r^{3B}$ && $ \left | \langle M \rangle - 3m - E_r^{3B} \right |$ \\ \hline
 Relativistic && 28.3876 && -35.9716 && -7.5840 && -7.5852 && 0.0012 \\ \hline
  && $\langle H_0 \rangle$ &&  $\langle V_{nr} \rangle$ && $\langle H \rangle$ && $E_{nr}^{3B}$ && $ \left | \langle H \rangle - E_{nr}^{3B} \right |$ \\ \hline
nonrelativistic && 29.7789 && -37.5173 && -7.7384 && -7.7380 && 0.0004  \\
 \hline
\end{tabular}
\end{center}
\caption{The expectation value of the 3B mass operator $\langle M \rangle$ obtained from free-body mass operator $\langle M_0 \rangle$ and the boost interaction $\langle V_k \rangle$ compared with the relativistic 3B binding energy $E_r^{3B}$. The corresponding nonrelativistic results are also shown. For both nonrelativistic and relativistic calculations, the MT-V potential is used. All the numbers are given in the units of MeV.}
\label{table:MT_expectation}
\end{table}

In Table \ref{table_Expectations} we present the contribution of different Faddeev components in the expectation value of the 3B mass operator $\bera \psi_i | M | \psi_j \ket - 3m $. To this aim the expectation values of the free-body mass operator $\bera \psi_i | M_0 | \psi_j \ket - 3m$ and the boost interaction $\bera \psi_i | V_k | \psi_j \ket $ are calculated for different Faddeev components. As one can see $\sum_{i,j=1}^3 \bera \psi_i | M_0 | \psi_j \ket - 3m = 28.3876\, \text{MeV}$, $\sum_{i,j=1}^3 \bera \psi_i | V_k | \psi_j \ket =-35.9716\, \text{MeV}$, and $ \sum_{i,j=1}^3 \bera \psi_i | M | \psi_j \ket - 3m =-7.5840\, \text{MeV}$.

\begin{table}[htb]
    \begin{tabularx}{\linewidth}{l*{3}{Y}}
    \toprule
    \multicolumn{4}{l}{\textbf{Panel A: $\bera \psi_i  | M_0 | \psi_j \ket - 3m$}} \\
    \midrule
  & $j=1$ & $j=2$ & $j=3$ \\ 
 \hline
 $i=1$ & 4.58426 & 2.43931 & 2.43931 \\ 
 $i=2$ & 2.43931 & 4.58396  & 2.43911 \\
 $i=3$ & 2.43931 &  2.43911 & 4.58396 \\
  \end{tabularx}
  \begin{tabularx}{\linewidth}{l*{3}{Y}}
    \toprule
    \multicolumn{4}{l}{\textbf{Panel B: $\bera \psi_i  | V_k | \psi_j \ket $}} \\
    \midrule
 $i=1$ & -6.88350 & -4.87431      & -4.87431 \\ 
 $i=2$ &  -4.87427 & -2.54511  & -2.25037      \\
 $i=3$ & -4.87427 &  -2.25037 & -2.54511      \\
    \bottomrule
   \multicolumn{4}{l}{\textbf{Panel C:  $\bera \psi_i  | M | \psi_j \ket  - 3m $}} \\
    \midrule
 $i=1$ & -2.29924 & -2.43500 & -2.43500 \\ 
 $i=2$ & -2.43496 & 2.03885  & 0.18873 \\
 $i=3$ & -2.43496 &  0.18873 & 2.03885 \\
    \bottomrule    
  \end{tabularx}
 \caption{The contribution of different Faddeev components in the expectation value of the 3B mass operator. Panel A: the expectation values of the free mass operator $\bera \psi_i  | M_0 | \psi_j \ket - 3m$, Panel B: the expectation values of the boost interaction $\bera \psi_i  | V_k | \psi_j \ket $, Panel C: the expectation values of the 3B mass operator $\bera \psi_i  | M | \psi_j \ket  - 3m $, calculated for different Faddeev components. All the numbers are given in the units of MeV.}
    \label{table_Expectations}
\end{table}

\section{Summary} \label{summary}
The inputs for the kernel of Faddeev and Yakubovsky integral equations, to study the relativistic corrections in three- and four-body bound and scattering states, are the boost 2B $t-$matrices. A number of theoretical methods are developed to calculate the relativistic and boost 2B $t-$matrices.
The direct solution of the relativistic Lippmann-Schwinger equation using relativistic interactions is one of the methods which has been successfully implemented in traditional partial wave decomposition calculations. One of the novel techniques for the calculation of the relativistic and boost interactions from the nonrelativistic interactions is solving a quadratic equation using an iterative scheme proposed by Kamada and Gl\"ockle.
In this paper, we apply the direct calculation of the boost interaction from the nonrelativistic interaction by solving a quadratic equation in a three-dimensional approach. To speed up the convergence in the solution of the quadratic equation, we use an averaging scheme. Then, by solving the Lippmann-Schwinger integral equation with the input boost 2B interaction, the boost 2B $t-$matrices are obtained as a function of the 2B relative momentum vectors, mainly the magnitude of the initial and final relative momenta and the angle between them. The 2B boost interactions have been then embedded into the kernel of three-dimensional Faddeev integral equations to study the relativistic corrections in the 3B binding energy. For a comprehensive study of the relativistic effects, one must consider four different relativistic corrections: the relativistic free propagator, the relativistic 2B $t-$matrices, the Jacobian, and the permutation coefficient. The contribution of each relativistic correction is studied in detail. We show that the relativistic corrections lead to a reduction of about $2\%$ in the 3B binding energy.
As a test for the numerical accuracy of our calculations, we calculate the expectation value of the 3B mass operator, which is in excellent agreement with the relativistic 3B binding energy. It is shown that in the nonrelativistic limit, where the momenta are much smaller than the masses, the nonrelativistic results are being reproduced from the relativistic ones.

\bibliography{references}

\section*{Acknowledgements}
The work of M. R. Hadizadeh was supported by the National Science Foundation under Grant Number NSF-HRD-1436702 with Central State University. M. Radin acknowledges partial financial support from Iran National Science Foundation (INSF).

\end{document}